\documentclass[11pt]{article}
\pdfoutput=1
\usepackage{dcolumn}
\usepackage{bm}
\usepackage{graphicx}
\usepackage{amssymb,amsmath}
\usepackage{multirow}
\usepackage{cite,color,url}
\usepackage[colorlinks=true
,urlcolor=blue
,anchorcolor=blue
,citecolor=blue
,filecolor=blue
,linkcolor=blue
,menucolor=blue
,linktocpage=true
,pdfproducer=medialab
,pdfa=true
]{hyperref}

\usepackage{slashed}
\usepackage{epsfig,psfrag,rotating,soul}
\usepackage{rotfloat}
\usepackage[font={small}]{caption}
\usepackage{colortbl}

\evensidemargin \oddsidemargin
\allowdisplaybreaks

\let\OLDthebibliography\thebibliography
\renewcommand\thebibliography[1]{
  \OLDthebibliography{#1}
  \setlength{\parskip}{0pt}
  \setlength{\itemsep}{0pt plus 0.3ex}
}

\definecolor{mygray}{gray}{0.85} 
\definecolor{myblue}{cmyk}{0.65, 0.37, 0.0, 0.19}

\begin{document}
\thispagestyle{empty}

\def\thefootnote{\fnsymbol{footnote}}

\begin{flushright}
IFT-UAM/CSIC-149
\end{flushright}

\vspace*{1cm}

\begin{center}

\begin{Large}
\textbf{\textsc{Hunting Squarks in Higgsino LSP scenarios at the LHC}}
\end{Large}

\vspace{1cm}

{\sc
Ernesto~Arganda$^{1, 2}$%
\footnote{{\tt \href{mailto:ernesto.arganda@csic.es}{ernesto.arganda@csic.es}}}%
, Antonio~Delgado$^{3}$%
\footnote{{\tt \href{mailto:adelgad2@nd.edu}{adelgad2@nd.edu}}}%
, Roberto~A.~Morales$^{2}$%
\footnote{{\tt \href{mailto:roberto.morales@fisica.unlp.edu.ar}{roberto.morales@fisica.unlp.edu.ar}}}%
,		 Mariano~Quir\'os$^{4}$%
\footnote{{\tt \href{quiros@ifae.es}{quiros@ifae.es}}}%

}

\vspace*{.7cm}

{\sl
$^1$Instituto de F\'{\i}sica Te\'orica UAM-CSIC, \\
C/ Nicol\'as Cabrera 13-15, Campus de Cantoblanco, 28049, Madrid, Spain

\vspace*{0.1cm}

$^2$IFLP, CONICET - Dpto. de F\'{\i}sica, Universidad Nacional de La Plata, \\ 
C.C. 67, 1900 La Plata, Argentina

\vspace*{0.1cm}

$^3$Department of Physics, University of Notre Dame, 225 Nieuwland Hall \\
Notre Dame, IN 46556, USA

\vspace*{0.1cm}

$^4$Institut de F\'{\i}sica d'Altes Energies (IFAE) and BIST, Campus UAB \\
08193, Bellaterra, Barcelona, Spain

}

\end{center}

\vspace{0.1cm}

\begin{abstract}
\noindent
The assumption that strongly interacting sparticles will decay directly to the LSP plus jets breaks down in situations where those decays are Yukawa suppressed. That occurs when producing the first two generations of squarks and when, at the same time, there are several electroweakinos lighter than those squarks. In this paper we analyze the signal of pair production of squarks that subsequently decay to an intermediate neutralino ($\tilde\chi_3^0$) plus jets. The neutralino will then decay to the LSP (mainly higgsino) and a Higgs. We have simulated the events and designed a discovery strategy based on a signal of two jets, four $b$-quarks and missing transverse energy. We obtain very promising values for the LHC sensitivity at 14 TeV and 300 fb$^{-1}$.
\end{abstract}

\def\thefootnote{\arabic{footnote}}
\setcounter{page}{0}
\setcounter{footnote}{0}

\newpage

\section{Introduction}
\label{intro}

Supersymmetry (SUSY) is one of the prime scenarios for physics beyond the Standard Model (SM). It can achieve a solution to the big hierarchy problem, accommodate gauge coupling unification and, if one assumes R-parity conservation, produce a candidate for Dark Matter. One of the consequences of R-parity conservation is that the lightest supersymmetric particle (LSP) is neutral and stable at the cosmological level, which means a lot of missing transverse energy (MET) on cascade decays at the Large Hadron Collider (LHC). In particular the minimal supersymmetric SM extension (MSSM) is, at present, the subject of extensive experimental searches at the LHC, mainly concerning gauginos, higgsinos (the supersymmetric partners of gauge and Higgs bosons). squarks and sleptons (the supersymmetric partners of quarks and leptons). 

In the MSSM the gaugino sector contains, on top of the gluino (a Majorana fermion), the electroweak neutralinos (four Majorana fermions $\tilde\chi_{i}^0$, $i=1,\dots 4$, an admixture of the bino, the neutral wino and the two neutral higgsinos) and charginos (two Dirac fermions $\tilde\chi_{a}^\pm$, $a=1,2$, an admixture of the two charged winos and the two charged higgsinos). The mass matrices for the neutralino and chargino sectors only depend, leaving aside the electroweak breaking parameters in the MSSM ($v$ and $\tan\beta$) on three mass parameters: the soft breaking masses for binos ($M_1$) and winos ($M_2$) and the supersymmetric higgsino mass ($\mu$). In view of the above comments it should be clear that the neutralino sector is pretty much independent on the particular mechanism of SUSY breaking in the scalar sector, as well as on the subsequently induced electroweak breaking. Moreover, in R-parity conserving scenarios, the neutralino sector is an excellent candidate to accommodate the LSP, i.e.~$\tilde\chi_1^0$, which makes experimental SUSY searches with lot of MET as model independent as possible. In this paper we will assume that the LSP is the lightest neutralino, and is mostly higgsino thus easily avoiding cosmological problems.

However in most of experimental analyses it is further assumed that the LSP $(\tilde\chi^0_1$) is an isolated neutral state and squark searches assume 100\% branching ratios (BR) to $\tilde\chi_1^0$ plus jet. This scenario is only possible, for neutralino masses larger than the electroweak scale, if the neutralino LSP is mostly bino. That assumption can have problems when one makes dark matter considerations, as the bino tends to overclose the universe~\cite{Arkani-Hamed:2006wnf}. If, on the other hand, one assumes that the higgsino is lighter that the bino and the wino,  since the higgsino is an $SU(2)$ doublet, one has a neutral state as LSP $(\tilde\chi^0_1)$ and a charged and a neutral one $(\tilde\chi^+_1, \tilde\chi^0_2)$  very close in mass. In this case the higgsino has no cosmological problems for masses below 1.1 TeV and, if there are no other electroweakinos with masses below the squark masses, the experimental signature is captured in current analyses that assume a 100\% BR to the LSP. Then even though there may be decays from squarks to the other two light electroweakinos $(\tilde\chi^+_1, \tilde\chi^0_2)$, subsequent decays of these electroweakinos will lead to the same signal since any of their decay products will be very soft.

A far more interesting situation, which we will consider in this paper, occurs when there is an intermediate neutral state $(\tilde\chi^3_0)$ between the squarks and the LSP (and associated states). This situation happens when there is a small hierarchy between the masses $(\mu<M_1<M_{\tilde q}<M_2)$, i.e.~the higgsino is lighter than the bino which in turn is lighter than the squarks, while the wino is essentially decoupled. This spectrum with $\tilde\chi_3^0$ lighter than strongly interacting sparticles was previously analyzed for the case of gluino production~\cite{Arganda:2021lpg, Arganda:2021iyr}, and also for the case of stop-pair production~\cite{Guchait:2021tmh}.

In this paper we are interested in decays of squarks within the MSSM for the case of higgsinos being the LSP, much heavier winos and first and second generations squarks at an intermediate scale between that of the bino and wino masses.  
We then assume the decay channel $\tilde q \to  \tilde\chi_3^0 j$, and prevent the direct decay into the LSP by assuming third generation squarks, that would decay into the LSP by the Yukawa interaction, heavier than first and second generation ones. The assumption that the third generation squarks are heavier than those of the first two generations, a hierarchy which appears naturally in the quark and lepton sectors in the SM, can be theoretically implemented in the effective theory of superstring models~\cite{Brignole:1997wnc}.

The paper is structured as follows: In Section~\ref{collider} we present a search strategy for this class of experimental signature at the LHC, characterizing the signal against the background. Section~\ref{results} is devoted to the analysis of the corresponding LHC sensitivity, by means of our search strategy, within the masses of the bino and squark parameter space. We finally conclude with a brief discussion of the main results in Section~\ref{conclus}.

\section{Squark-pair Production at the LHC: Collider Study}
\label{collider}

\begin{figure}[ht!]
	\begin{center}
		\begin{tabular}{c}
			\centering
			\hspace*{-3mm}
			\includegraphics[scale=0.75]{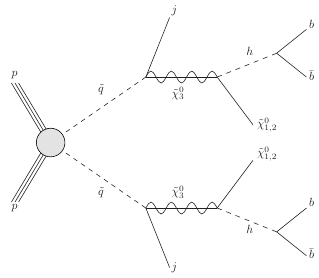} 
		\end{tabular}
		\caption{\it Relevant SUSY process corresponding to squark production with $2 j+ 4 b + E_T^\text{miss}$ in the final state.}
		\label{fig:process}
	\end{center}
\end{figure}

Within the MSSM higgsino LSP scenarios presented above, 
the proposed experimental LHC signature originates from the squark-pair production, $pp \to \tilde q \tilde q^*$, that decay into $\tilde \chi_3^0$ (essentially a bino) and one light jet ($\tilde q \to  \tilde\chi_3^0 j$). Each $\tilde\chi_3^0$ decays then into the LSP ($\tilde\chi_1^0$) and $h$, the lightest MSSM Higgs boson considered here as the 125-GeV Higgs boson, which in turn decays into a $b$-quark pair. Thus, the LHC signature is made of two light jets, four $b$-jets, and a lot of missing transverse energy ($2 j+ 4 b + E_T^\text{miss}$), as shown in Fig.~\ref{fig:process}. The main backgrounds are separated into the following categories: QCD multijet; $t \bar t$ production; $t \bar t$ production associated with electroweak or Higgs bosons, $t \bar t$ + $X$ ($X$ = $W$, $Z$, $\gamma^*$, $h$); $Z$ + jets and $W$ + jets productions; and diboson production ($WW$, $ZZ$, $WZ$, $Wh$, and $Zh$) plus jets.

Our LHC search strategy is developed for a center-of-mass energy of $\sqrt{s}$ = 14 TeV and an integrated luminosity of $\cal{L}$ = 1 ab$^{-1}$, corresponding to the high-luminosity LHC (HL-LHC) stage. {\tt MadGraph\_aMC@NLO 2.8.1}~\cite{Alwall:2014hca} is used for the Monte Carlo generation of the signal and the main backgrounds, the parton showering and hadronization are obtained with {\tt PYTHIA 8.2}~\cite{Sjostrand:2014zea}, and the detector response is simulated with {\tt Delphes 3.3.3}~\cite{deFavereau:2013fsa}. 

On the other hand, we have confronted our spectrum of interest with the general searches at the LHC of new physics for 13 TeV by using the software {\tt CheckMATE 2.0.24}~\cite{Dercks:2016npn} and we conclude that the spectrum is allowed by the validated analyses. In particular, the resulting $r$ parameters are below 0.1 and the most sensitive search is (see Ref.~\cite{ATLAS:2017vat}) in the $SRI-MLL-60$ region.
From this comparison, the general validated searches in {\tt CheckMATE} are far from being sensitive to squark-pair production with our spectrum, so that it is relevant to develop a dedicated search strategy for the signature considered here. Notice that as the $r$ parameter values are too low, we do not expect that the current experimental analysis for ${\cal L}$=1000 fb$^{-1}$ exclude our spectrum.

For the simulation of the signal, the default cuts on the transverse momenta of the light jets and $b$-jets are used ($p_T^j >$ 20 GeV and $p_T^b >$ 20 GeV).
However, in order to decrease the large background cross sections and make event generation more efficient, the generator-level cuts of Eq.~(\ref{generatorcuts}), on the $p_T$ of the light jets and $b$-jets, for the background simulation are imposed, since very energetic $b$-jets and light jets from the decays of Higgs bosons and squarks are expected, respectively. 
\begin{eqnarray}
&p_T^{j_1} > 180 \, \text{GeV} \,, \quad &p_T^{j_2} > 140 \, \text{GeV} \,, \quad p_T^{j_3} > 70 \, \text{GeV} \,, \quad p_T^{j_4} > 35 \, \text{GeV} \,, \nonumber\\
&p_T^{b_1} > 90 \, \text{GeV} \,, \quad &p_T^{b_2} > 20 \, \text{GeV} \,, \quad p_T^{b_3} > 20 \, \text{GeV} \,, \quad p_T^{b_4} > 20 \, \text{GeV} \,,
\label{generatorcuts}
\end{eqnarray}
with $j_1$ ($b_1$) being the most energetic light ($b$-) jet and $j_4$ ($b_4$) the least one. 
The MLM algorithm~\cite{Mangano:2002,Mangano:2006rw} has been implemented for jet matching and merging. With the intention of optimizing the event simulation and checking that the jet distributions are smooth, {\tt xqcut} is chosen to be 20 for all generated samples and {\tt qcut} equal to 30, 50, and 250 for $t \bar t$, backgrounds with bosons, and signal, respectively. A working point for the efficiency of $b$-tagging of 0.75 is used, with a rate of misidentification of $0.01$ for light jets and $0.1$ for $c$-jets (internal analysis codes and simulation input files are available upon request to authors).	

It is appropriate to make the following comments on the signal and its corresponding backgrounds:
\begin{itemize}
\item Signal. Supersymmetric spectrum and decay rates for the signal have been calculated with {\tt SOFTSUSY.4.1.10}~\cite{Allanach:2001kg,Allanach:2017hcf,Allanach:2013kza,Allanach:2009bv,Allanach:2011de,Allanach:2014nba,Allanach:2016rxd}, while the production cross section of a pair of squarks is obtained from~\cite{LHCSUSYxs}. We consider two cases: the \textit{Left} case, for $\tilde{u}_L\tilde{u}_L$ and $\tilde{d}_L\tilde{d}_L$ productions, and the \textit{Right} case, for $\tilde{u}_R\tilde{u}_R$ production. In both cases, we keep the LSP mass almost fixed around 500 GeV, and we scan over the parameter space $M_{\tilde q}\in[800,1100]$ GeV and $M_{\tilde\chi_3^0}\in[600,900]$ GeV. 
In these parameter space regions, for the \textit{Left} case we have $\tilde{q}_L\tilde{q}_L$ production cross-sections in the range $[45,160]$ fb and BR($\tilde{q}_L\to q\,\tilde\chi_3^0$)$\in[0.19,0.88]$.
On the other hand, for the \textit{Right} case, we have $\sigma(p\,p\to\tilde{u}_R\tilde{u}_R)\in[58,400]$ fb and BR($\tilde{u}_R\to u\,\tilde\chi_3^0$)$\in[0.56,0.95]$. However, production cross-sections for $p\,p\to\tilde{d}_R\tilde{d}_R$ are typically one order of magnitude smaller than $\tilde{q}_L\tilde{q}_L$ and $\tilde{u}_R\tilde{u}_R$ cases and we will not take this process into account (see for instance Ref.~\cite{Hollik:2012rc}).
\item QCD multijet background. It is commonly treated using data-driven techniques and is intractable with our computational capacity. An estimate of this background by recasting the analysis in~\cite{ATLAS:2020syg} is included, in which a similar cut-based analysis is performed for a similar experimental signature. In particular, as our signal is expected to have a large amount of $E_T^\text{miss}$, variables related to this observable, such as the $E_T^\text{miss}$ significance, will substantially decrease this class of backgrounds with non-genuine missing transverse energy. Moreover, the characteristic $\vec{p}_T^\text{\,miss}$ spatial configuration can be used for reducing this background.
\item $t \bar{t}$ production. Its fully-hadronic and semileptonic decay channels are included. The branching fraction of the fully-hadronic channel is 0.457 while that of the semileptonic is 0.438. After applying the cuts of Eq.~(\ref{generatorcuts}), one expects 1.36 $\times 10^6$ events for the fully-hadronic channel and 0.42 $\times 10^6$ events for the semileptonic one. To be somewhat more realistic with the simulation of this background, we consider an extra jet, which translates to 0.83 $\times 10^6$ and 0.25 $\times 10^6$ more events for the fully-hadronic and semileptonic channels, respectively. An estimate of $t \bar{t}+2j$ is also included through an extra factor of 10\% for the simulated events of $t \bar{t}$ plus $t \bar{t}+j$ (this 10\% comes from the ratio of the corresponding cross sections).
\item $t\bar{t} + X$ production. In relation to $t\bar{t}$, the extra boson generates a genuine source of $E_T^\text{miss}$ (more $b$-jets) for the hadronic (semileptonic) top-quark pair. $t\bar{t}_{\rm semilep}+(h\to b\bar{b})$, $t\bar{t}_{\rm semilep}+(\gamma^*\to b\bar{b})$,  $t\bar{t}_{\rm had}+(W\to l\nu)$, $t\bar{t}_{\rm had}+(Z\to\nu\nu)$, and $t\bar{t}_{\rm semilep}+(Z\to b\bar{b})$ are considered with one extra jet to each process, which leads to $2.9\times 10^3$ expected events in this category.
\item $V$+jets production ($V =$ $Z$ or $W$). A $b$-jet pair and a light-jet pair leading to 4 extra jets and a genuine source of $E_T^\text{miss}$ through neutrinos coming from the gauge boson decays (with BR($Z\to\nu\nu$) = 0.2 and BR($W\to l\nu)$ = 0.21). $5.6 \times 10^4$ events for $Z$+jets and $3 \times 10^5$ events for $W$+jets are expectable for $\cal{L}$ = 1000 fb$^{-1}$.
\item Diboson production. It is subdominant, with $\sim 10^{-3}$ times the $V$+jets number of events (which we will keep under control). Therefore, one can safely neglect this background.
\end{itemize}
In what follows we will characterize the signal against the dominant SM backgrounds, which will allow us to find signal regions suitable for our search strategy.
In the collider study to be developed, the backgrounds we have just defined fall into four categories: $t \bar{t}_{\rm had}+2j$ and $t \bar{t}_{\rm semilep}+2j$ (both inclusive), $V$+jets, and $t \bar{t}+X+j$ (also inclusive).

\begin{figure}[ht!]
	\begin{center}
		\begin{tabular}{cc}
			\centering
			\hspace*{-3mm}
			\includegraphics[scale=0.4]{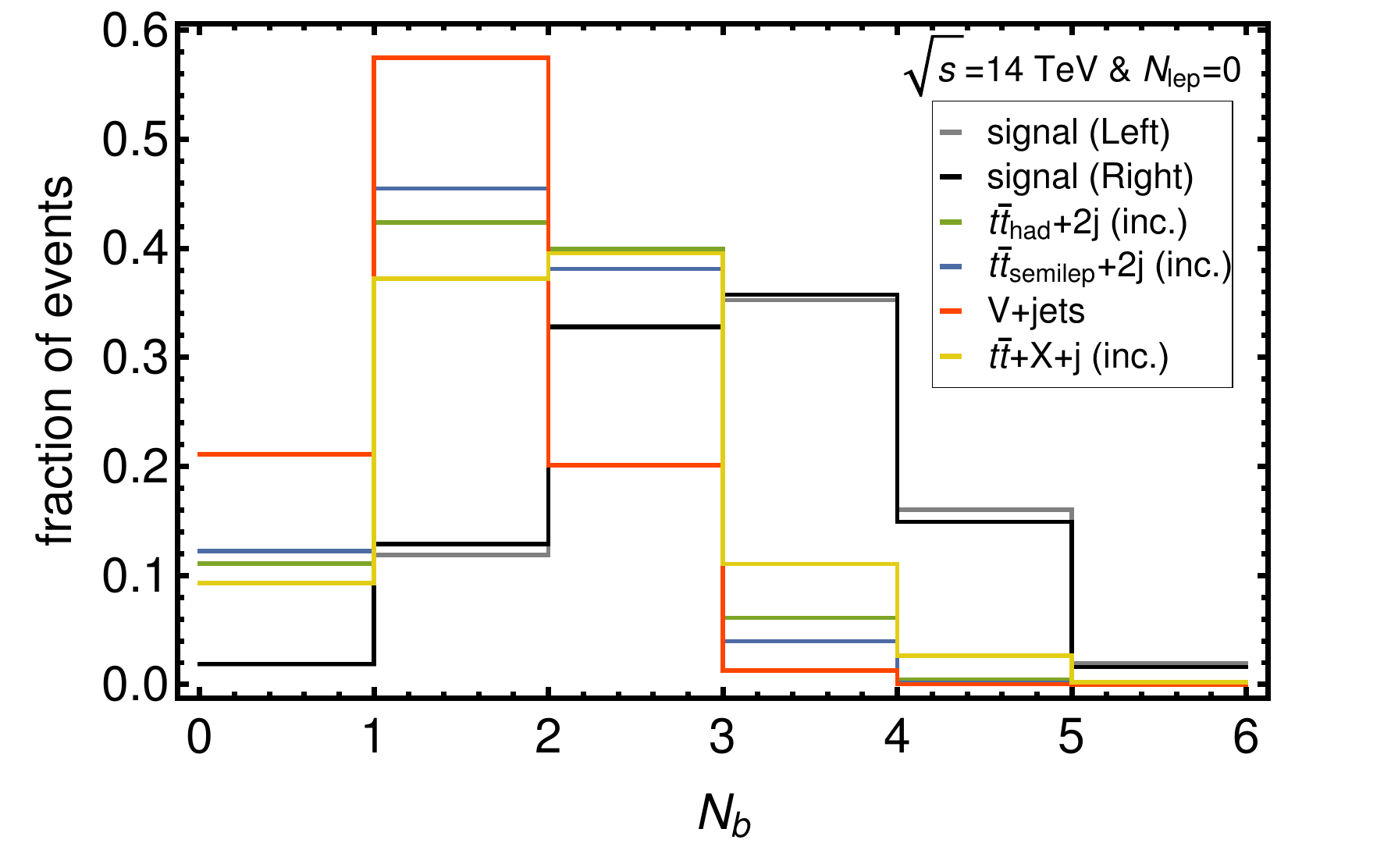} &
			\includegraphics[scale=0.4]{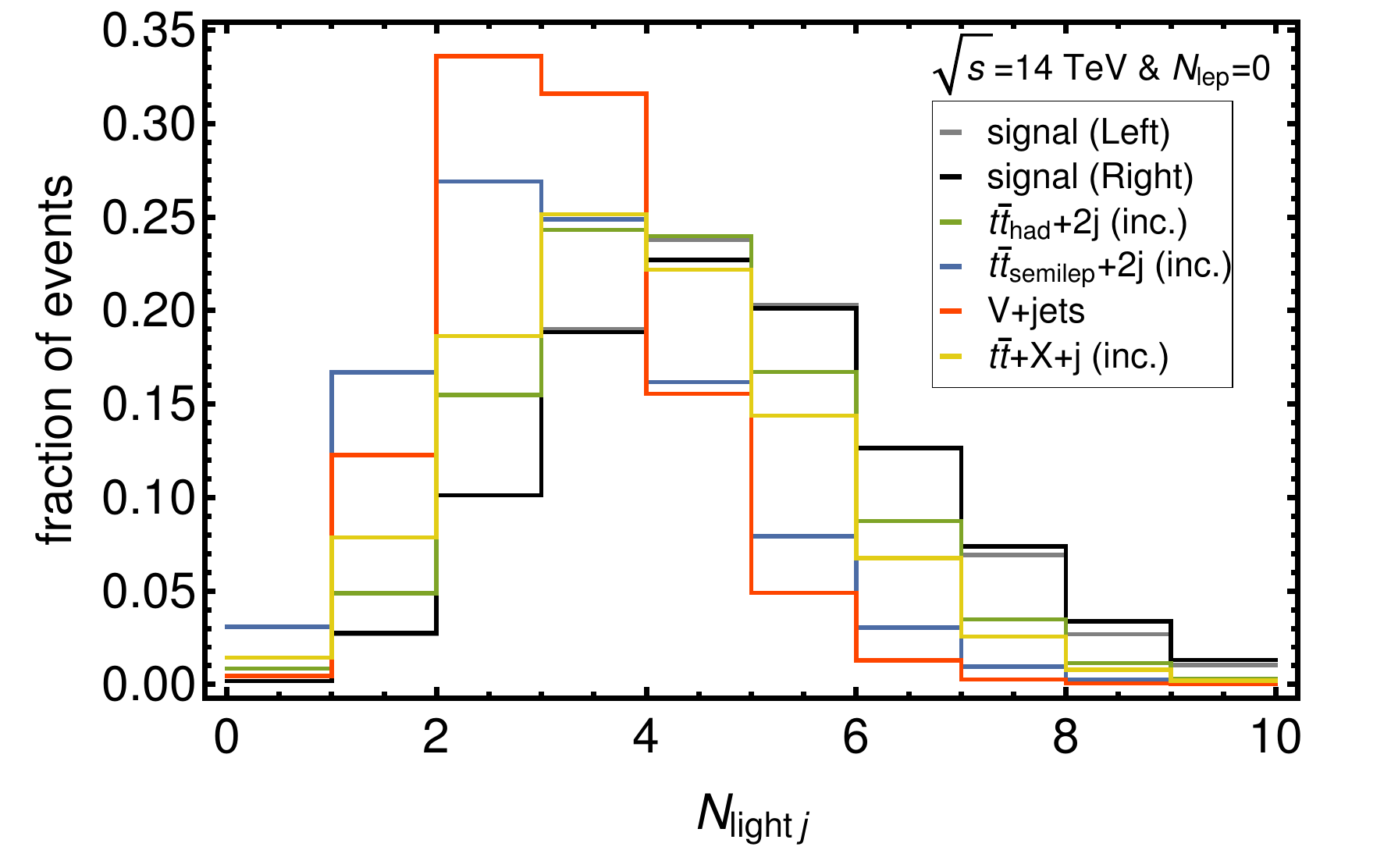}	
		\end{tabular}
		\caption{\it Distributions (with $N_\ell =$ 0) of the fraction of $N_b$ (left panel) and $N_j$ (right panel) for signal and background.}
		\label{fig:Nb-Nj}
	\end{center}
\end{figure}

Distributions of the fraction of the number of identified $b$-jets $N_b$ and the number of light jets $N_j$ for the signal and the main backgrounds are depicted in Fig.~\ref{fig:Nb-Nj} (left and right panels, respectively). We consider the generic squark masses of 1 TeV for both \textit{Left} and \textit{Right} production cases. First, a lepton veto ($N_\ell$ = 0) is set (already imposed on Fig.~\ref{fig:Nb-Nj}) in order to reduce the semileptonic $t \bar t$ production, which is one of the most dangerous backgrounds. It is important to note that the maximum of the $N_j$ distribution for the signal is found for 4 light jets, while most of the backgrounds are chopped for a lower value of the light-jet multiplicity. In addition, the fact of having four bottom quarks coming from the two Higgs-boson decays is a challenging task, since in principle one wants to identify the four $b$-jets.
All of this motivates, on the one hand, the requirement of at least four light jets in our search strategy and, on the other hand, the definition of two signal regions: a first signal region (SR1) requiring at least four $b$-jets in the final state, and a second one (SR2) with at least three identified $b$-jets.

\begin{figure}[ht!]
	\begin{center}
		\begin{tabular}{cc}
			\centering
			\hspace*{-3mm}
			\includegraphics[scale=0.4]{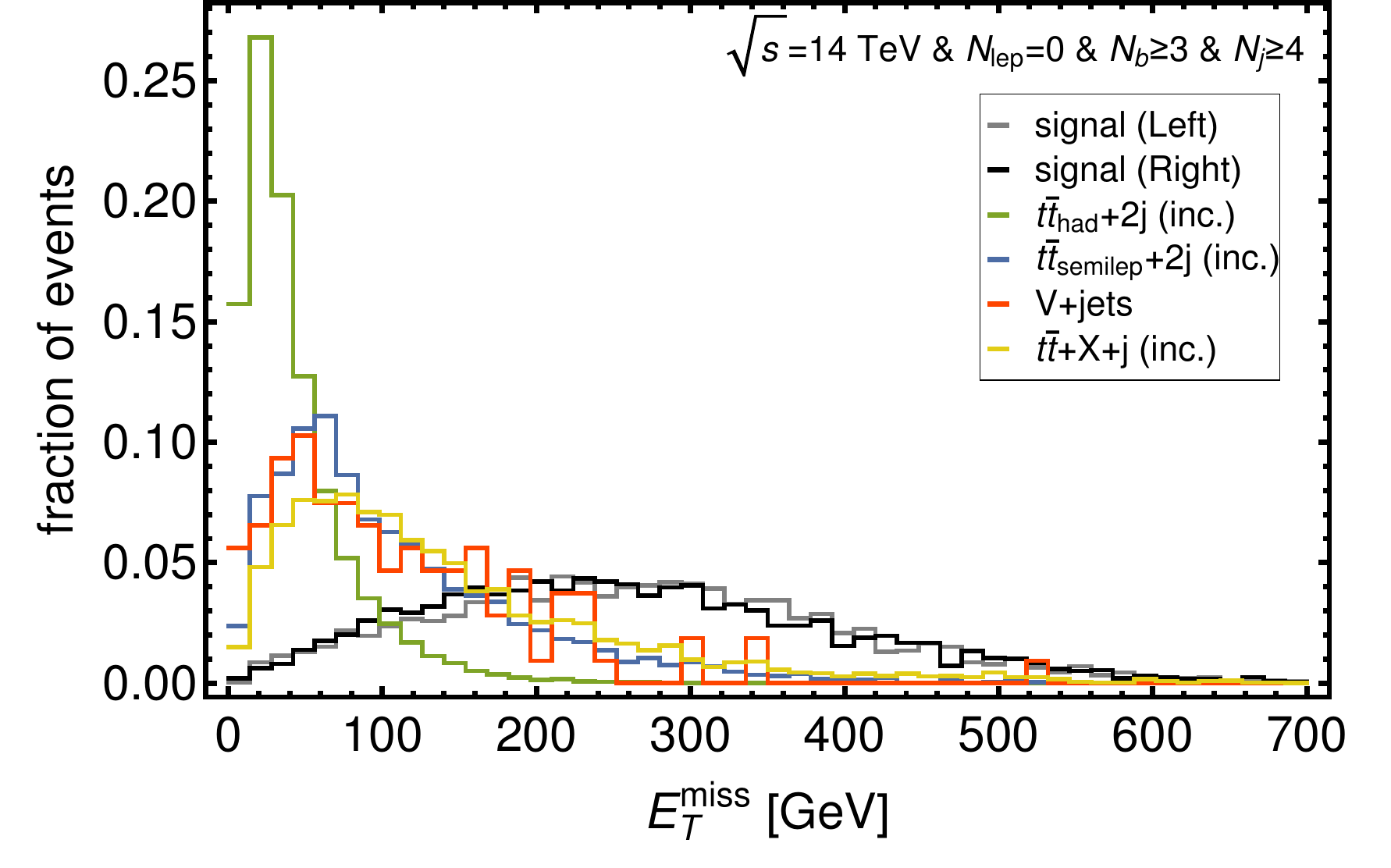} &			\includegraphics[scale=0.4]{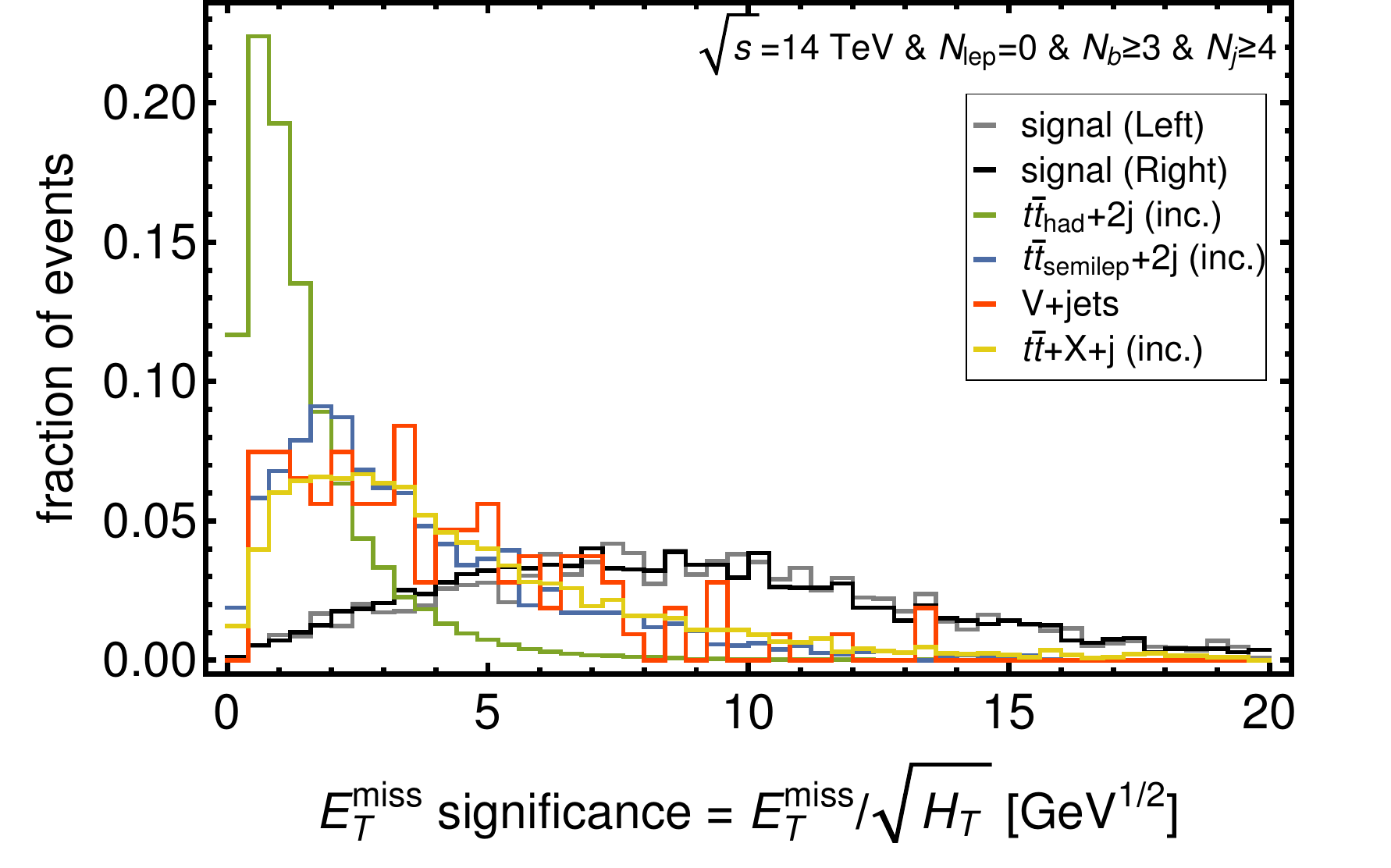} \\
			\includegraphics[scale=0.4]{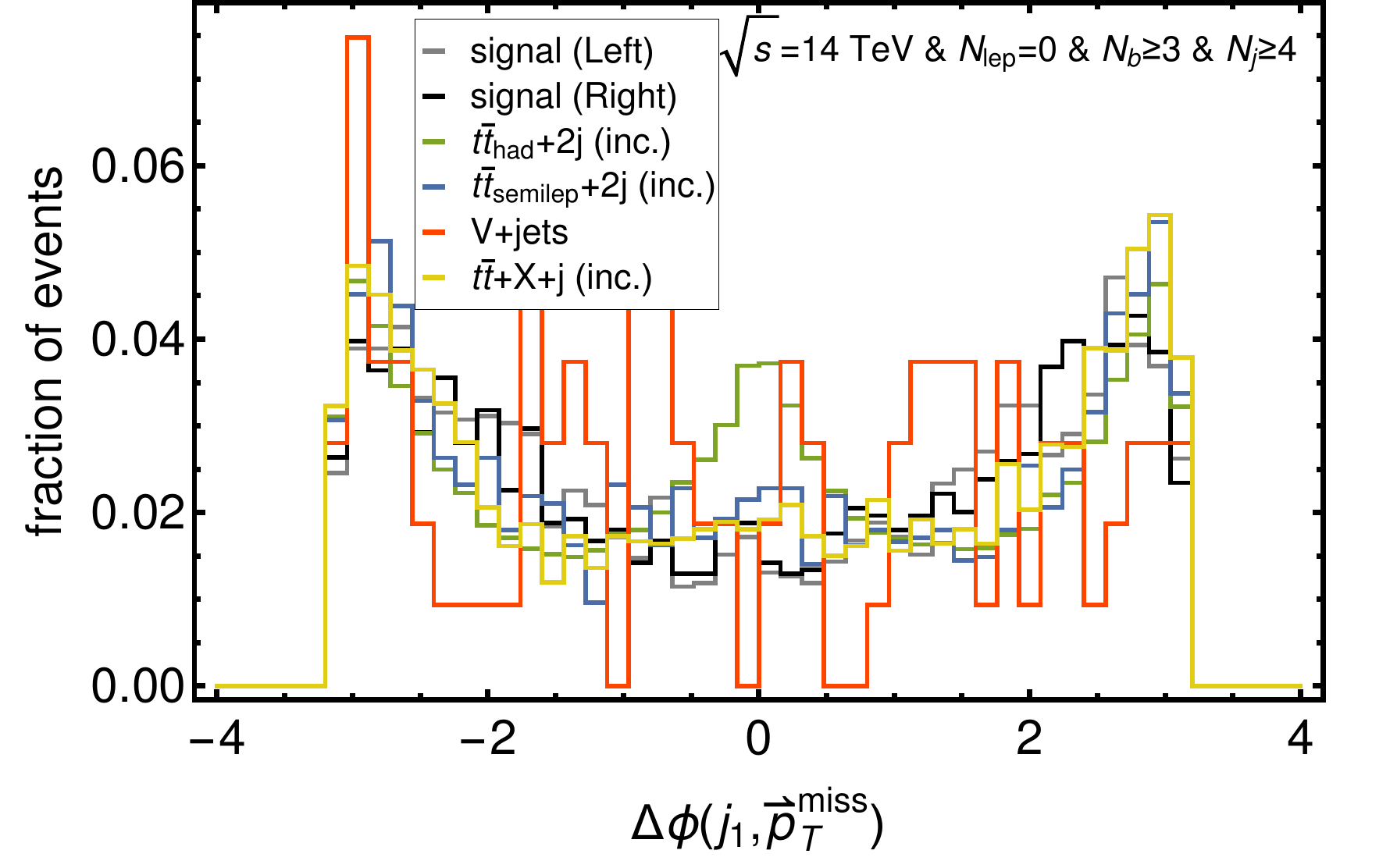} & \includegraphics[scale=0.4]{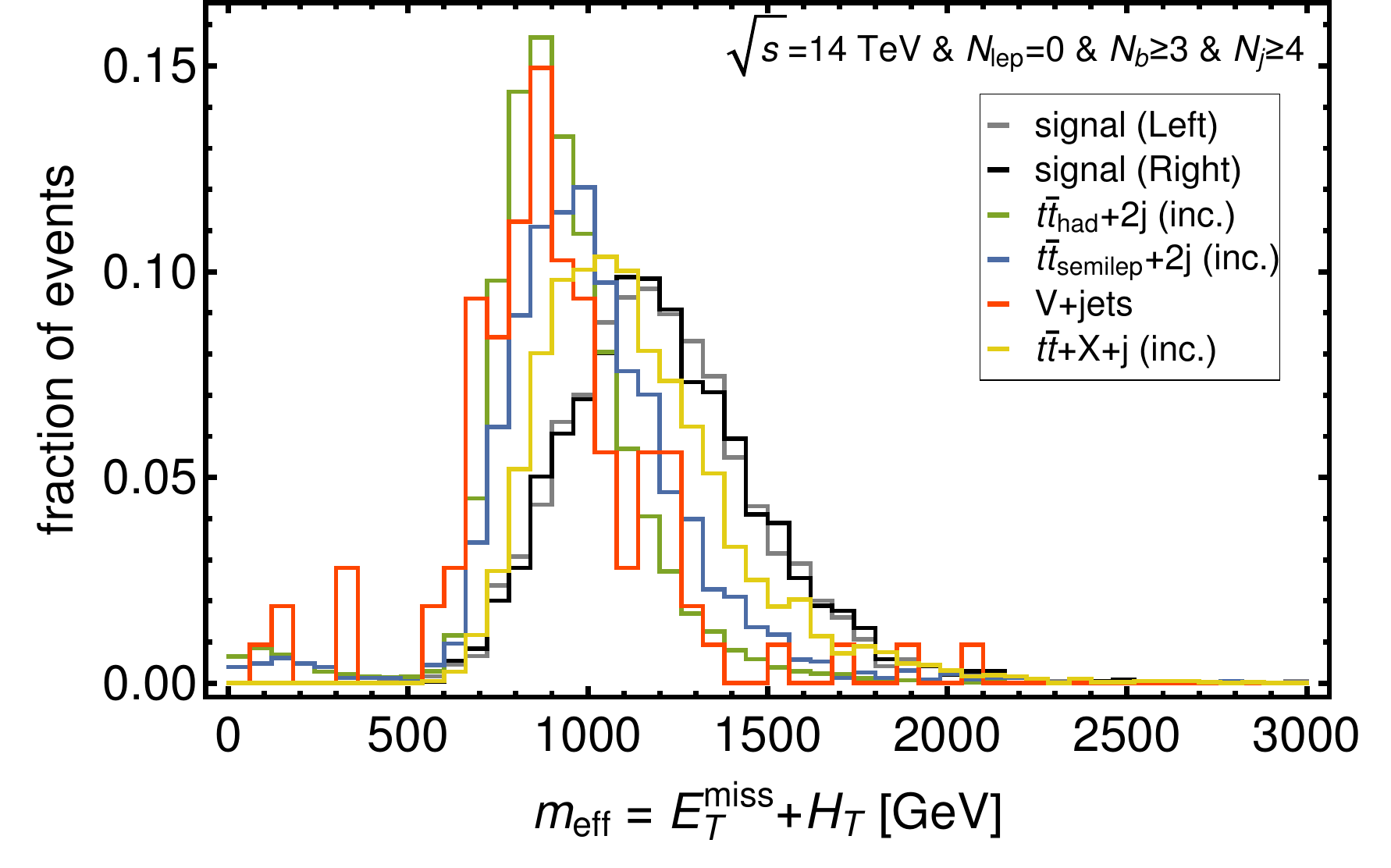}
		\end{tabular}
		\caption{\it Distributions (with $N_\ell =$ 0, at least 3 $b$-jets and 4 light jets) of the fraction of $E_T^\text{miss}$ (upper left panel), the $E_T^\text{miss}$ significance (upper right panel), the azimuthal angle difference $\Delta\phi$($j_1$, $\vec{p}_T^\text{\,miss})$ between the leading jet and the $\vec{p}_T^\text{\,miss}$ (lower left panel), and the effective mass $m_\text{eff}$ (lower right panel), for signal and background.}
		\label{fig:MET-dist}
	\end{center}
\end{figure}

Fig.~\ref{fig:MET-dist} is devoted to the usual distributions, in particular, those related to the missing transverse energy $E_T^\text{miss}$. 
The missing energy distributions for the backgrounds have peaks below 100 GeV while the signal have most of its events above this value.

Furthermore, the $E_T^\text{miss}$ significance distributions of signal and background present different patterns with a clear separation near 5 GeV$^{1/2}$.
The $\Delta\phi$($j_1$,$\vec{p}_T^\text{\,miss})$ distributions show a non-central behavior for signal and most of the irreducible backgrounds but this variable is useful in order to reject most of the QCD multijet background.
Following the data-driven analysis of this background in \cite{ATLAS:2020syg}, we will include a conservative estimate of 0.7 events as final number for the multijet background in the cut flow for ${\cal L}$=1000 fb$^{-1}$ (corresponding to a direct extrapolation from ${\cal L}=139$ fb$^{-1}$ to the luminosity considered here in the signal regions with at least 4 jets).

Finally, the $m_\text{eff}$ variable is constructed as the sum of the missing transverse energy and the scalar sum of the transverse momenta of all the reconstructed jets ($H_T$). The corresponding distributions of the background events have peaks below 1000 GeV, while the signal peak is near this value.

Taking all this into account, the definition of the SR1 search strategy contains the cuts listed below:
\begin{itemize}
\item selection cuts $N_b \geq 4 \,,\,\, N_j \geq 4 \,,\,\, N_\ell = 0$,
\item $\chi_{HH}$~\footnote{Following the definition in~\cite{ATLAS:2018rnh}, this variable quantifies the requirement of two reconstructed bottom pairs coming from the Higgs bosons.} cut lower than 2,
\item MET cuts given by $E_T^\text{miss}>$ 150 GeV, $\vert\Delta\phi$($j_1$, $\vec{p}_T^\text{\,miss})\vert>0.4$, and $m_\text{eff}>$ 1300 GeV,
\end{itemize}
while the search strategy for the signal region SR2 is defined with the following cuts:
\begin{itemize}
\item selection cuts $N_b \geq 3 \,,\,\, N_j \geq 4 \,,\,\, N_\ell = 0$,
\item $p_T$ cuts of Eq.~(\ref{pTcuts}),
\begin{eqnarray}
&&p_T^{j_1} > 200 \, \text{GeV} \,, \quad p_T^{j_2} > 150 \, \text{GeV} \,, \quad p_T^{j_3} > 80 \, \text{GeV} \,, \quad p_T^{j_4} > 40 \, \text{GeV} \,, \nonumber\\
&&p_T^{b_1} > 100 \, \text{GeV} \,, \quad p_T^{b_2} > 60 \, \text{GeV} \,, \quad p_T^{b_3} > 35 \, \text{GeV} \,.
\label{pTcuts}
\end{eqnarray}
\item MET cuts given by $E_T^\text{miss}>$ 150 GeV, $\vert\Delta\phi$($j_1$, $\vec{p}_T^\text{\,miss})\vert>0.4$, and $m_\text{eff}>$ 1400 GeV.
\end{itemize}

For the statistical analysis, we consider the significance including background systematic uncertainties~\cite{Cowan:2010js,Cowan:2012} given by:
\begin{equation}
{\cal S} = \sqrt{2 \left((B+S) \log \left(\frac{(S+B)(B+\sigma_{B}^{2})}{B^{2}+(S+B)\sigma_{B}^{2}}\right)-\frac{B^{2}}{\sigma_{B}^{2}}\log \left(1+\frac{\sigma_{B}^{2}S}{B(B+\sigma_{B}^{2})} \right) \right)} \,,
\label{systS}
\end{equation}
where $S$ ($B$) is the number of signal (background) events and $\sigma_{B}=(\Delta B) B$, with $\Delta B$ being the relative systematic uncertainty chosen to be a conservative value of 30\%. This conservative value takes the uncertainty associated to our limited statistics into account, since we mostly suppress the expected backgrounds by means of the use of our search strategy, as shown in the next section.


\section{Results}
\label{results}

The cut-by-cut resulting significances with systematic uncertainties of 30\%, with our search strategy applied to each signal region (SR1, SR2) and to \textit{Left}, \textit{Right} productions, are summarized along Tables \ref{cutflowSR1LEFT}-\ref{cutflowSR2RIGHT}. 
Remember that the search strategy was developed for $\cal{L}$ = 1000 fb$^{-1}$, and a QCD multijet estimate of 0.7 events~\cite{ATLAS:2020syg} was included in the significances of the last cut.

As we can see, most of the backgrounds were eliminated and we keep at least 5 signal events at the end of the strategies.
In particular, the results for the SR1 in which we demand four reconstructed $b$-jets, corresponding to \textit{Left} and \textit{Right} productions, are collected in Tables \ref{cutflowSR1LEFT} and \ref{cutflowSR1RIGHT}, respectively. 
For each kind of signal, we expect significances of 3.16$\sigma$ and 2.74$\sigma$ for $\cal{L}$ = 1000 fb$^{-1}$. Even results at $\cal{L}$ = 300 fb$^{-1}$ are interesting with significances near 2$\sigma$.

\begin{table}
\hspace*{-12.5mm}
    \centering
\begin{tabular}{r|rrrrr|cc}
\hline\hline
   Process  & signal & $t\bar{t}_{\rm had}+2j$ (inc.) & $t\bar{t}_{\rm semilep}+2j$ (inc.) & $V$+jets & $t\bar{t}X+j$ (inc.) & $\cal{S}$ \\
   \hline
    Expected  & 2110 & $2.4 \times 10^6$ & $0.74 \times 10^6$ & $3.56 \times 10^5$ & $2.9\times 10^3$ & $2\times 10^{-3}$ \\
    \hline
    selection cuts  & 173.1 & 2697 & 90.2 & 7.95 & 13.2 & $1.3\times 10^{-2}$ \\
    $\chi_{HH}$ cut  & 38.6 & 364.8 & 12.7 & 0 & 1.7 & 0.32 \\
    MET and $m_\text{eff}$ cuts & 6.9 & 1.1 & 0 & 0 & 0.1 & 3.16 \\
    \hline\hline
    $\cal{L}$ = 300 fb$^{-1}$& 2.1 & 0.3 & 0 & 0 & 0 & 1.89 \\
    \hline\hline
\end{tabular}
    \caption{\it Cut flow for SR1 with $\cal{L}$ = 1000 fb$^{-1}$ for the \textit{Left} case signal production. Significances from Eq.~(\ref{systS}), with a systematic uncertainty in the background of 30\%. A QCD multijet estimate of 0.7 events~\cite{ATLAS:2020syg} is taken into account for the significances of the last cut-flow step.}
    \label{cutflowSR1LEFT}
\end{table}

\begin{table}
\hspace*{-12.5mm}
    \centering
\begin{tabular}{r|rrrrr|cc}
\hline\hline
   Process  & signal & $t\bar{t}_{\rm had}+2j$ (inc.) & $t\bar{t}_{\rm semilep}+2j$ (inc.) & $V$+jets & $t\bar{t}X+j$ (inc.) & $\cal{S}$ \\
   \hline
    Expected  & 1701 & $2.4 \times 10^6$ & $0.74 \times 10^6$ & $3.56 \times 10^5$ & $2.9\times 10^3$ & $1.6\times 10^{-3}$ \\
    \hline
    selection cuts  & 136.7 & 2697 & 90.2 & 7.95 & 13.2 & 0.16 \\
    $\chi_{HH}$ cut  & 31.9 & 364.8 & 12.7 & 0 & 1.7 & 0.27 \\
    MET and $m_\text{eff}$ cuts & 5.7 & 1.1 & 0 & 0 & 0.1 & 2.74 \\
    \hline\hline
    $\cal{L}$ = 300 fb$^{-1}$& 1.72 & 0.3 & 0 & 0 & 0 & 1.63 \\
    \hline\hline
\end{tabular}
    \caption{\it Cut flow for SR1 with $\cal{L}$ = 1000 fb$^{-1}$ for the \textit{Right} case signal production. Significances from Eq.~(\ref{systS}), with a systematic uncertainty in the background of 30\%. A QCD multijet estimate of 0.7 events~\cite{ATLAS:2020syg} is taken into account for the significances of the last cut-flow step.}
    \label{cutflowSR1RIGHT}
\end{table}

On the other hand, Tables \ref{cutflowSR2LEFT} and \ref{cutflowSR2RIGHT} show the results for \textit{Left} and \textit{Right} productions in the SR2 (in which we demand three reconstructed $b$-jets).
Then, for this signal region, the results are very promising with significances above 4$\sigma$ (7$\sigma$) for luminosity of 300 (1000) fb$^{-1}$.

\begin{table}
\hspace*{-12.5mm}
    \centering
\begin{tabular}{r|rrrrr|cc}
\hline\hline
   Process  & signal & $t\bar{t}_{\rm had}+2j$ (inc.) & $t\bar{t}_{\rm semilep}+2j$ (inc.) & $V$+jets & $t\bar{t}X+j$ (inc.) & $\cal{S}$ \\
   \hline
    Expected  & 2110 & $2.4 \times 10^6$ & $0.74 \times 10^6$ & $3.56 \times 10^5$ & $2.9\times 10^3$ & $2\times 10^{-3}$ \\
    \hline
    selection cuts  & 616.9 & $3.06 \times 10^4$ & 2025 & 145.7 & 94.1 & 0.06 \\
    $p_T$ cuts  & 35.9 & 216.7 & 4.1 & 0 & 2.1 & 0.49 \\
    MET and $m_\text{eff}$ cuts & 16.8 & 0 & 0.3 & 0 & 0. & 7.22 \\
    \hline\hline
    $\cal{L}$ = 300 fb$^{-1}$& 5 & 0 & 0.1 & 0 & 0 & 4.32 \\
    \hline\hline
\end{tabular}
    \caption{\it Cut flow for SR2 with $\cal{L}$ = 1000 fb$^{-1}$  for the \textit{Left} case signal production. Significances from Eq.~(\ref{systS}), with a systematic uncertainty in the background of 30\%. A QCD multijet estimate of 0.7 events~\cite{ATLAS:2020syg} is taken into account for the significances of the last cut-flow step.}
    \label{cutflowSR2LEFT}
\end{table}

\begin{table}
\hspace*{-12.5mm}
    \centering
\begin{tabular}{r|rrrrr|cc}
\hline\hline
   Process  & signal & $t\bar{t}_{\rm had}+2j$ (inc.) & $t\bar{t}_{\rm semilep}+2j$ (inc.) & $V$+jets & $t\bar{t}X+j$ (inc.) & $\cal{S}$ \\
   \hline
    Expected  & 1701 & $2.4 \times 10^6$ & $0.74 \times 10^6$ & $3.56 \times 10^5$ & $2.9\times 10^3$ & $1.6\times 10^{-3}$ \\
    \hline
    selection cuts  & 508.9 & $3.06 \times 10^4$ & 2025 & 145.7 & 94.1 & 0.05 \\
    $p_T$ cuts  & 38.3 & 216.7 & 4.1 & 0 & 2.1 & 0.05 \\
    MET and $m_\text{eff}$ cuts & 18.1 & 0 & 0.3 & 0 & 0. & 7.58 \\
    \hline\hline
    $\cal{L}$ = 300 fb$^{-1}$& 5.4 & 0 & 0.1 & 0 & 0 & 4.55 \\
    \hline\hline
\end{tabular}
    \caption{\it Cut flow for SR2 with $\cal{L}$ = 1000 fb$^{-1}$ for the \textit{Right} case signal production. Significances from Eq.~(\ref{systS}), with a systematic uncertainty in the background of 30\%. A QCD multijet estimate of 0.7 events~\cite{ATLAS:2020syg} is taken into account for the significances of the last cut-flow step.}
    \label{cutflowSR2RIGHT}
\end{table}

\begin{figure}
	\begin{center}
		\begin{tabular}{cc}
			\centering
			\hspace*{-3mm}
			\includegraphics[scale=0.9]{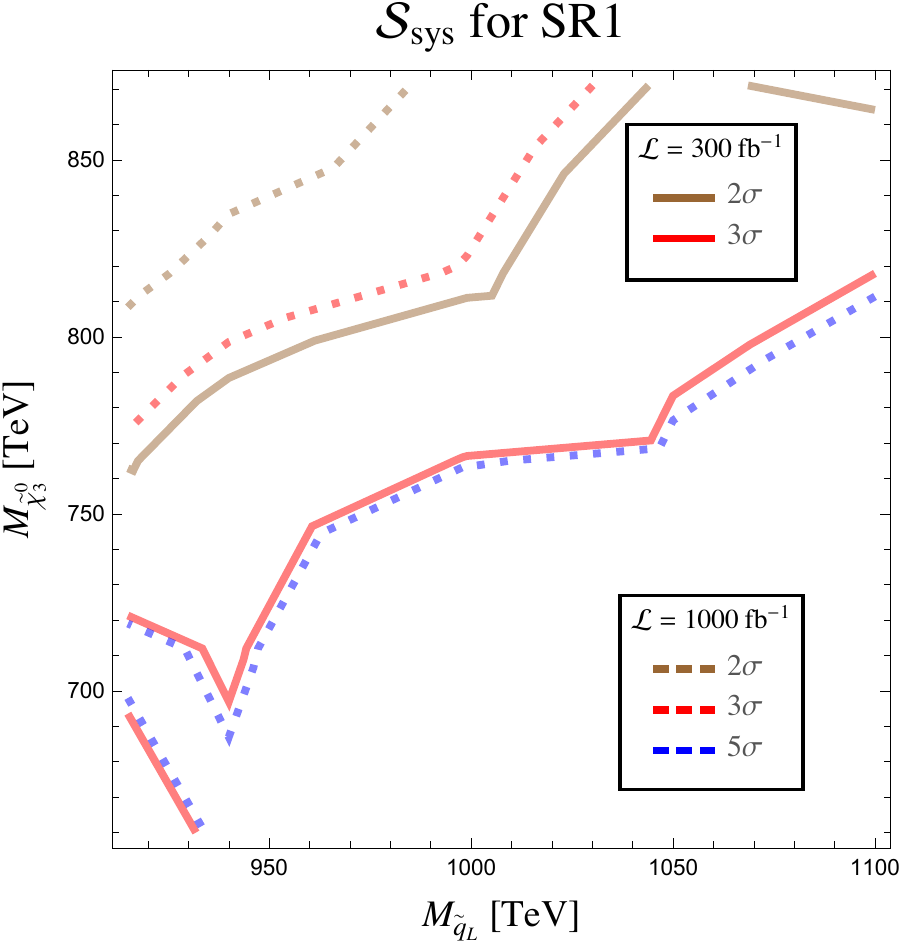} &
			\includegraphics[scale=0.9]{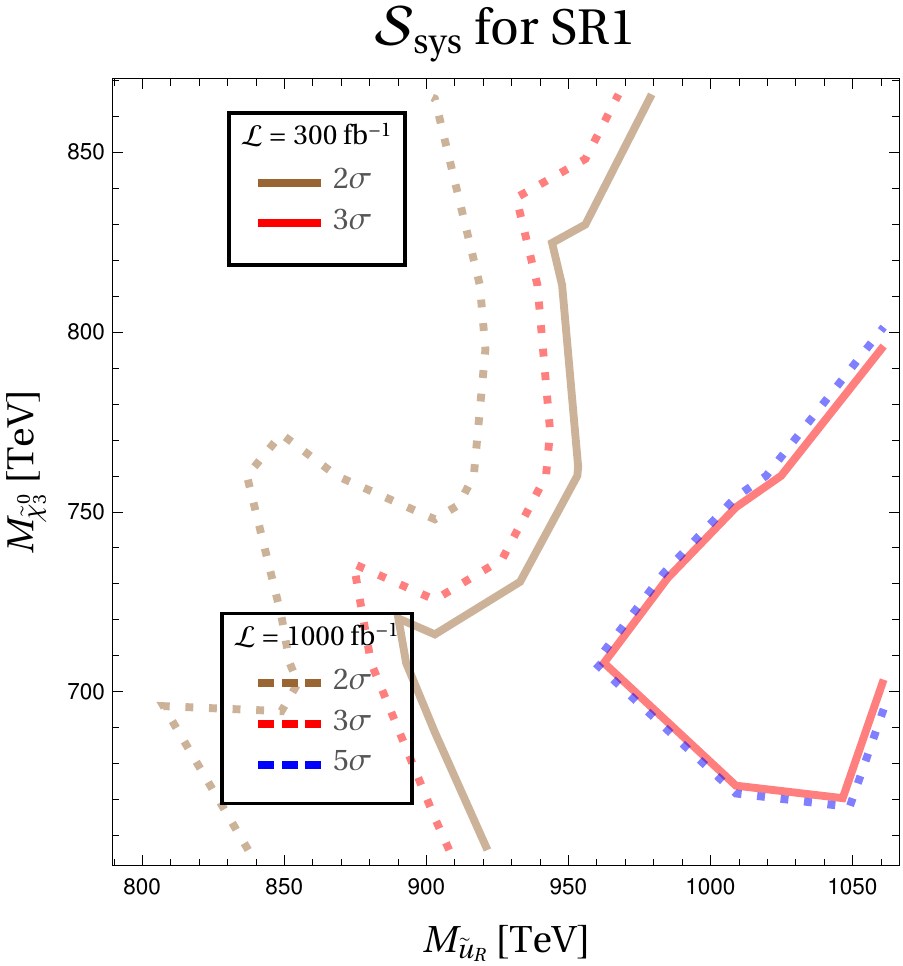}	
		\end{tabular}
		\caption{\it Contour lines for SR1 in the plane $[M_{\tilde q},M_{\tilde\chi_3^0}]$ corresponding to $\tilde{q}_L$ (left) and $\tilde{u}_R$ (right) productions. The brown, red and blue colors are the $\cal{S}$ (background systematic uncertainty of 30\%) with values of 2$\sigma$, 3$\sigma$ and 5$\sigma$, respectively. Solid (dotted) lines correspond to ${\cal L}$ = 300 (1000) fb$^{-1}$.}
		\label{contourplotsSR1}
	\end{center}
\end{figure}

The search strategy exploits the large amount of $E_T^\text{miss}$ and $p_T$ of the several energetic $b$ and light jets, which are very characteristic for the spectrum that we are considering, i.e.~with the higgsino LSP coming from an intermediate bino state.
Hence, we applied this general search strategy in order to study the sensitivity in the plane [$M_{\tilde q}$, $M_{\tilde\chi_3^0}$], corresponding to the relevant parameters of these SUSY scenarios in which we fix the LSP mass to 500 GeV and decouple the rest of the spectrum. In particular, we explored the ranges of interest of $M_{\tilde q}\in[800,1100]$ GeV and $M_{\tilde\chi_3^0}\in[600,900]$ GeV.
The contour lines of $\cal{S}$ for SR1 corresponding to \textit{Left} and \textit{Right} productions are shown in Fig.~\ref{contourplotsSR1}. The solid and dashed lines correspond to ${\cal L}$ = 300 and 1000 fb$^{-1}$, respectively, and the brown, red, blue lines correspond to values of $\cal{S}$ of 2$\sigma$, 3$\sigma$, and 5$\sigma$, respectively. 
With a luminosity of 300 fb$^{-1}$, we obtain significances at the evidence level in most of the region for $M_{\tilde{q}_L}\gtrsim 950$ GeV and $M_{\tilde\chi_3^0}\lesssim 750$ GeV in the \textit{Left} case. However, this region is reduced in the \textit{Right} case. Notice that the projections to ${\cal L}$ = 1000 fb$^{-1}$ with discovery-level significance are very similar to the previous ones in both cases.

\begin{figure}
	\begin{center}
		\begin{tabular}{cc}
			\centering
			\hspace*{-3mm}
			\includegraphics[scale=0.9]{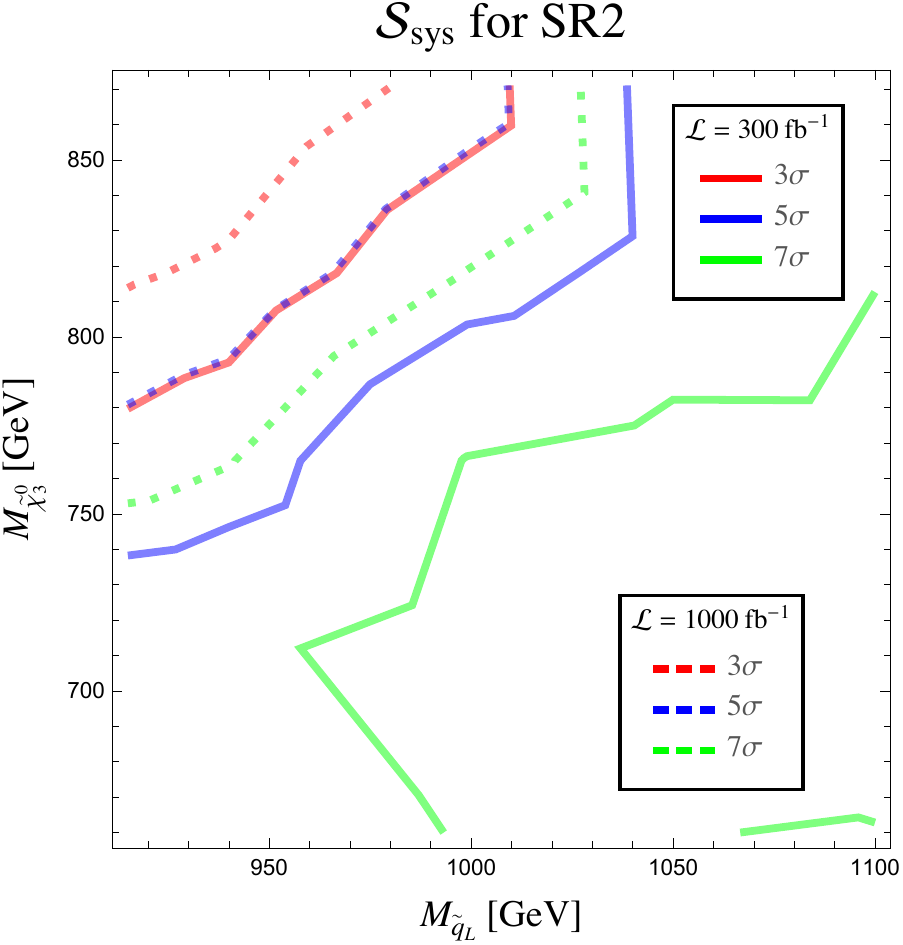} &
			\includegraphics[scale=0.9]{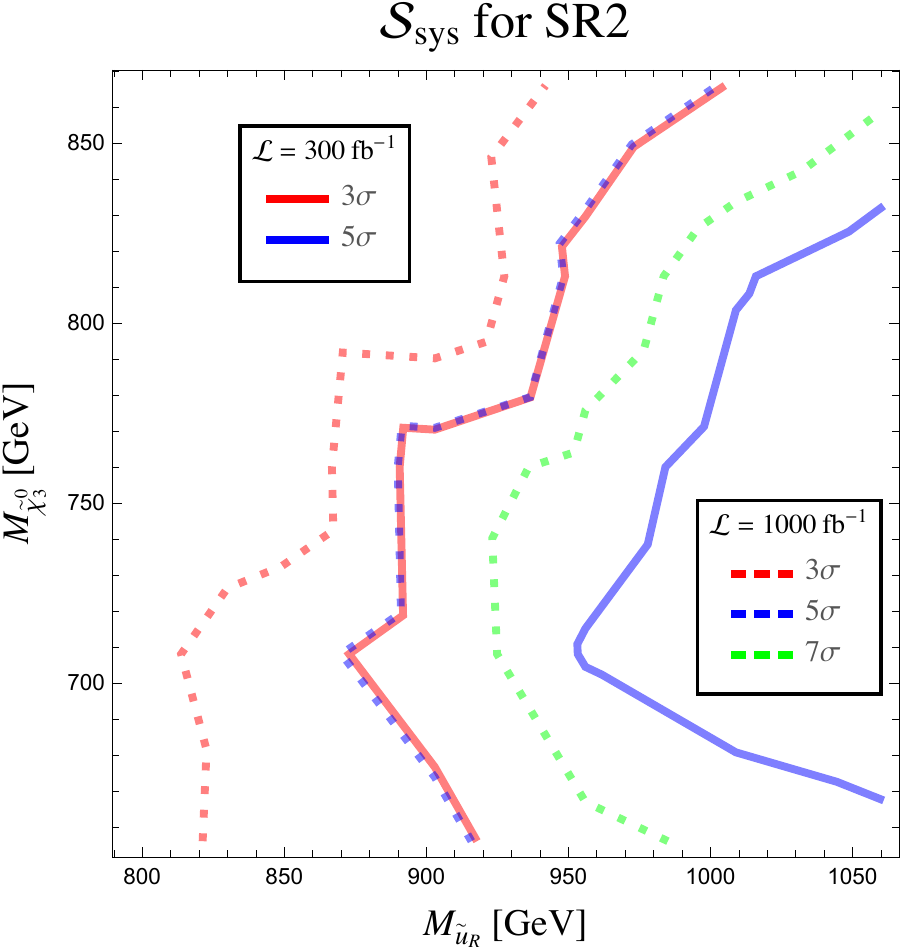}	
		\end{tabular}
		\caption{\it Contour lines for SR2 in the plane $[M_{\tilde q},M_{\tilde\chi_3^0}]$ corresponding to $\tilde{q}_L$ (left) and $\tilde{u}_R$ (right) productions. The red, blue and green colors are the $\cal{S}$ (background systematic uncertainty of 30\%) with values of 3$\sigma$, 5$\sigma$ and 7$\sigma$, respectively. Solid (dotted) lines correspond to ${\cal L}$ = 300 (1000) fb$^{-1}$.}
		\label{contourplotsSR2}
	\end{center}
\end{figure}

For the SR2 in Fig.~\ref{contourplotsSR2} we also include the green lines corresponding to significances of 7$\sigma$ since the results are improved with respect to the SR1 ones. We expect more signal events after the cuts of this search strategy, and discovery-level significances arise even for ${\cal L}$ = 300 fb$^{-1}$. In particular, we have $\cal{S}\geq$ 5$\sigma$ for $M_{\tilde{q}_L}\gtrsim 950$ GeV and $M_{\tilde\chi_3^0}\lesssim 750$ GeV in the \textit{Left} case, and $M_{\tilde{q}_R}\gtrsim$ 1000 GeV and 700 GeV $\lesssim M_{\tilde\chi_3^0}\lesssim$ 800 GeV for the \textit{Right} case. It is not superfluous to again reiterate here that the parameter space of interest of these SUSY scenarios is not excluded by any of the experimental searches carried out at the LHC by the ATLAS and CMS collaborations, as we have checked using {\tt CheckMATE}. Therefore, with the very promising results obtained for LHC sensitivity, with both search strategies, we think that it would be interesting to develop experimental analyses of these channels starting from the signal interpretation that we propose.


\section{Brief discussion}
\label{conclus}

Some comments are in order to summarize the main conclusions of this work. 
When one deviates from the assumption of a 100\% branching fractions of supersymmetric decays to the LSP, the current searches have a different interpretation. In this paper we have analyzed a particular spectrum such that the mass of the first two squark generations is heavier than a mostly bino neutralino ($\tilde\chi_0^3$), which in turn is heavier than a mostly higgisino set of states ($\tilde\chi^0_{1,2}$, $\tilde\chi^+_1$). Pair production of squarks at the LHC will generate a cascade decay via $\tilde\chi_3^0$, since the direct decay to the LSP is Yukawa suppressed. 

We have first shown that, for the values of the masses considered along this work, the current LHC searches provide no bounds. Then we have studied the feasibility of discovery of this signal with two light jets, four $b$-quarks and MET. The results are very promising, specially in the SR2 case where one can obtain even 7$\sigma$ significances for 300 fb$^{-1}$, within regions of the parameter space of the doublet case. The general message we want to convey is that one needs to deviate from simplified models to be able to cover a wide range of the parameter space in this class of BSM scenarios.

\section*{Acknowledgments}
The work of EA is partially supported by the ``Atracci\'on de Talento'' program (Modalidad 1) of the Comunidad de Madrid (Spain) under the grant number 2019-T1/TIC-14019, by the Spanish Research Agency (Agencia Estatal de Investigaci\'on) through the grant IFT Centro de Excelencia Severo Ochoa SEV-2016-0597, and by CONICET and ANPCyT (Argentina) under projects PICT 2016-0164, PICT 2017-2751, and PICT 2017-2765. The work of AD was partially supported by the National Science Foundation under grant PHY-2112540. The work of RM was supported by CONICET (Argentina). The work of MQ is partly supported by Spanish MINEICO under Grant FPA2017-88915-P, by the Catalan Government under Grant 2017SGR1069, and by Severo Ochoa Excellence Program of MINEICO under Grant SEV-2016-0588. IFAE is partially funded by the CERCA program of the Generalitat de Catalunya.

\bibliographystyle{JHEP}
\bibliography{lit.bib}

\providecommand{\href}[2]{#2}\begingroup\raggedright\begin{thebibliography}{10}

\bibitem{Arkani-Hamed:2006wnf}
N.~Arkani-Hamed, A.~Delgado and G.~F. Giudice, \emph{{The Well-tempered
  neutralino}},
  \href{https://doi.org/10.1016/j.nuclphysb.2006.02.010}{\emph{Nucl. Phys. B}
  {\bfseries 741} (2006) 108}
  [\href{https://arxiv.org/abs/hep-ph/0601041}{{\ttfamily hep-ph/0601041}}].

\bibitem{Arganda:2021lpg}
E.~Arganda, A.~Delgado, R.~A. Morales and M.~Quir\'os, \emph{{Novel Higgsino
  dark matter signal interpretation at the LHC}},
  \href{https://doi.org/10.1103/PhysRevD.104.055003}{\emph{Phys. Rev. D}
  {\bfseries 104} (2021) 055003}
  [\href{https://arxiv.org/abs/2104.13827}{{\ttfamily 2104.13827}}].

\bibitem{Arganda:2021iyr}
E.~Arganda, A.~Delgado, R.~A. Morales and M.~Quir\'os, \emph{{Search strategy
  for gluinos at the LHC with a Higgs boson decaying into tau leptons}},
  \href{https://arxiv.org/abs/2107.06034}{{\ttfamily 2107.06034}}.

\bibitem{Guchait:2021tmh}
M.~Guchait, A.~Roy and S.~Sharma, \emph{{Probing mild-tempered neutralino dark
  matter through top-squark production at the LHC}},
  \href{https://doi.org/10.1103/PhysRevD.104.055032}{\emph{Phys. Rev. D}
  {\bfseries 104} (2021) 055032}
  [\href{https://arxiv.org/abs/2103.09810}{{\ttfamily 2103.09810}}].

\bibitem{Brignole:1997wnc}
A.~Brignole, L.~E. Ibanez and C.~Munoz, \emph{{Soft supersymmetry breaking
  terms from supergravity and superstring models}},
  \href{https://doi.org/10.1142/9789812839657_0003}{\emph{Adv. Ser. Direct.
  High Energy Phys.} {\bfseries 18} (1998) 125}
  [\href{https://arxiv.org/abs/hep-ph/9707209}{{\ttfamily hep-ph/9707209}}].

\bibitem{Alwall:2014hca}
J.~Alwall, R.~Frederix, S.~Frixione, V.~Hirschi, F.~Maltoni, O.~Mattelaer
  et~al., \emph{{The automated computation of tree-level and next-to-leading
  order differential cross sections, and their matching to parton shower
  simulations}}, \href{https://doi.org/10.1007/JHEP07(2014)079}{\emph{JHEP}
  {\bfseries 07} (2014) 079} [\href{https://arxiv.org/abs/1405.0301}{{\ttfamily
  1405.0301}}].

\bibitem{Sjostrand:2014zea}
T.~Sjöstrand, S.~Ask, J.~R. Christiansen, R.~Corke, N.~Desai, P.~Ilten et~al.,
  \emph{{An Introduction to PYTHIA 8.2}},
  \href{https://doi.org/10.1016/j.cpc.2015.01.024}{\emph{Comput. Phys. Commun.}
  {\bfseries 191} (2015) 159}
  [\href{https://arxiv.org/abs/1410.3012}{{\ttfamily 1410.3012}}].

\bibitem{deFavereau:2013fsa}
{\scshape DELPHES 3} collaboration, \emph{{DELPHES 3, A modular framework for
  fast simulation of a generic collider experiment}},
  \href{https://doi.org/10.1007/JHEP02(2014)057}{\emph{JHEP} {\bfseries 02}
  (2014) 057} [\href{https://arxiv.org/abs/1307.6346}{{\ttfamily 1307.6346}}].

\bibitem{Dercks:2016npn}
D.~Dercks, N.~Desai, J.~S. Kim, K.~Rolbiecki, J.~Tattersall and T.~Weber,
  \emph{{CheckMATE 2: From the model to the limit}},
  \href{https://doi.org/10.1016/j.cpc.2017.08.021}{\emph{Comput. Phys. Commun.}
  {\bfseries 221} (2017) 383}
  [\href{https://arxiv.org/abs/1611.09856}{{\ttfamily 1611.09856}}].

\bibitem{ATLAS:2017vat}
{\scshape ATLAS} collaboration, \emph{{Search for electroweak production of
  supersymmetric states in scenarios with compressed mass spectra at
  $\sqrt{s}=13$ TeV with the ATLAS detector}},
  \href{https://doi.org/10.1103/PhysRevD.97.052010}{\emph{Phys. Rev. D}
  {\bfseries 97} (2018) 052010}
  [\href{https://arxiv.org/abs/1712.08119}{{\ttfamily 1712.08119}}].

\bibitem{Mangano:2002}
M.~Mangano, \emph{{The so-called MLM prescription for ME/PS matching}},
  {\emph{Fermilab ME/MC Tuning Workshop, October 4, 2002,
  \url{http://www-cpd.fnal.gov/personal/mrenna/tuning/nov2002/mlm.pdf.gz}
  \href{http://www-cpd.fnal.gov/personal/mrenna/tuning/nov2002/mlm.pdf.gz}}
  (2002) }.

\bibitem{Mangano:2006rw}
M.~L. Mangano, M.~Moretti, F.~Piccinini and M.~Treccani, \emph{{Matching matrix
  elements and shower evolution for top-quark production in hadronic
  collisions}},
  \href{https://doi.org/10.1088/1126-6708/2007/01/013}{\emph{JHEP} {\bfseries
  01} (2007) 013} [\href{https://arxiv.org/abs/hep-ph/0611129}{{\ttfamily
  hep-ph/0611129}}].

\bibitem{Allanach:2001kg}
B.~C. Allanach, \emph{{SOFTSUSY: a program for calculating supersymmetric
  spectra}}, \href{https://doi.org/10.1016/S0010-4655(01)00460-X}{\emph{Comput.
  Phys. Commun.} {\bfseries 143} (2002) 305}
  [\href{https://arxiv.org/abs/hep-ph/0104145}{{\ttfamily hep-ph/0104145}}].

\bibitem{Allanach:2017hcf}
B.~C. Allanach and T.~Cridge, \emph{{The Calculation of Sparticle and Higgs
  Decays in the Minimal and Next-to-Minimal Supersymmetric Standard Models:
  SOFTSUSY4.0}},  \href{https://arxiv.org/abs/1703.09717}{{\ttfamily
  1703.09717}}.

\bibitem{Allanach:2013kza}
B.~C. Allanach, P.~Athron, L.~C. Tunstall, A.~Voigt and A.~G. Williams,
  \emph{{Next-to-Minimal SOFTSUSY}},
  \href{https://doi.org/10.1016/j.cpc.2014.04.015}{\emph{Comput. Phys. Commun.}
  {\bfseries 185} (2014) 2322}
  [\href{https://arxiv.org/abs/1311.7659}{{\ttfamily 1311.7659}}].

\bibitem{Allanach:2009bv}
B.~C. Allanach and M.~A. Bernhardt, \emph{{Including R-parity violation in the
  numerical computation of the spectrum of the minimal supersymmetric standard
  model: SOFTSUSY}},
  \href{https://doi.org/10.1016/j.cpc.2009.09.015}{\emph{Comput. Phys. Commun.}
  {\bfseries 181} (2010) 232}
  [\href{https://arxiv.org/abs/0903.1805}{{\ttfamily 0903.1805}}].

\bibitem{Allanach:2011de}
B.~C. Allanach, C.~H. Kom and M.~Hanussek, \emph{{Computation of Neutrino
  Masses in R-parity Violating Supersymmetry: SOFTSUSY3.2}},
  \href{https://doi.org/10.1016/j.cpc.2011.11.024}{\emph{Comput. Phys. Commun.}
  {\bfseries 183} (2012) 785}
  [\href{https://arxiv.org/abs/1109.3735}{{\ttfamily 1109.3735}}].

\bibitem{Allanach:2014nba}
B.~C. Allanach, A.~Bednyakov and R.~Ruiz~de Austri, \emph{{Higher order
  corrections and unification in the minimal supersymmetric standard model:
  SOFTSUSY3.5}}, \href{https://doi.org/10.1016/j.cpc.2014.12.006}{\emph{Comput.
  Phys. Commun.} {\bfseries 189} (2015) 192}
  [\href{https://arxiv.org/abs/1407.6130}{{\ttfamily 1407.6130}}].

\bibitem{Allanach:2016rxd}
B.~C. Allanach, S.~P. Martin, D.~G. Robertson and R.~R. de~Austri, \emph{{The
  Inclusion of Two-Loop SUSYQCD Corrections to Gluino and Squark Pole Masses in
  the Minimal and Next-to-Minimal Supersymmetric Standard Model: SOFTSUSY3.7}},
   \href{https://arxiv.org/abs/1601.06657}{{\ttfamily 1601.06657}}.

\bibitem{LHCSUSYxs}
C.~Borschensky, Z.~Gecse, M.~Kraemer, R.~van~der Leeuw, A.~Kulesza, M.~Mangano
  et~al., ``{LHC SUSY Cross Section Working Group}.''
  \url{https://twiki.cern.ch/twiki/bin/view/LHCPhysics/SUSYCrossSections},
  2020.

\bibitem{Hollik:2012rc}
W.~Hollik, J.~M. Lindert and D.~Pagani, \emph{{NLO corrections to squark-squark
  production and decay at the LHC}},
  \href{https://doi.org/10.1007/JHEP03(2013)139}{\emph{JHEP} {\bfseries 03}
  (2013) 139} [\href{https://arxiv.org/abs/1207.1071}{{\ttfamily 1207.1071}}].

\bibitem{ATLAS:2020syg}
{\scshape ATLAS} collaboration, \emph{{Search for squarks and gluinos in final
  states with jets and missing transverse momentum using 139 fb$^{-1}$ of
  $\sqrt{s}$ =13 TeV $pp$ collision data with the ATLAS detector}},
  \href{https://doi.org/10.1007/JHEP02(2021)143}{\emph{JHEP} {\bfseries 02}
  (2021) 143} [\href{https://arxiv.org/abs/2010.14293}{{\ttfamily
  2010.14293}}].

\bibitem{ATLAS:2018rnh}
{\scshape ATLAS} collaboration, \emph{{Search for pair production of Higgs
  bosons in the $b\bar{b}b\bar{b}$ final state using proton-proton collisions
  at $\sqrt{s} = 13$ TeV with the ATLAS detector}},
  \href{https://doi.org/10.1007/JHEP01(2019)030}{\emph{JHEP} {\bfseries 01}
  (2019) 030} [\href{https://arxiv.org/abs/1804.06174}{{\ttfamily
  1804.06174}}].

\bibitem{Cowan:2010js}
G.~Cowan, K.~Cranmer, E.~Gross and O.~Vitells, \emph{{Asymptotic formulae for
  likelihood-based tests of new physics}},
  \href{https://doi.org/10.1140/epjc/s10052-011-1554-0}{\emph{Eur. Phys. J. C}
  {\bfseries 71} (2011) 1554}
  [\href{https://arxiv.org/abs/1007.1727}{{\ttfamily 1007.1727}}].

\bibitem{Cowan:2012}
G.~Cowan, \emph{{Discovery sensitivity for a counting experiment with
  background uncertainty}}, {\emph{tech. rep., Royal Holloway, London (2012)
  \url{http://www.pp.rhul.ac.uk/~cowan/stat/medsig/medsigNote.pdf}
  \href{http://www.pp.rhul.ac.uk/~cowan/stat/medsig/medsigNote.pdf}} (2012) }.

\end{thebibliography}\endgroup

\end{document}